\documentclass[3p,times,twocolumn]{elsarticle}

\usepackage{ecrc}

\volume{00}

\firstpage{1}

\journalname{Nuclear Physics B Proceedings Supplement}

\runauth{C.~D.~Froggatt, R.~Nevzorov, H.~B.~Nielsen, A.~W.~Thomas}

\jid{nuphbp}

\jnltitlelogo{Nuclear Physics B Proceedings Supplement}

\usepackage{amssymb}

\usepackage[figuresright]{rotating}

\begin{document}

\begin{frontmatter}



\dochead{}

\title{On the smallness of the cosmological constant}


\author[UoG]{C.~D.~Froggatt}

\author[UoA,ITEP]{R.~Nevzorov}

\author[NBI]{H.~B.~Nielsen}

\author[UoA]{A.~W.~Thomas}

\address[UoG]{School of Physics and Astronomy, University of Glasgow, Glasgow, UK}
\address[UoA]{ARC Centre of Excellence for Particle Physics at the Tera--scale,\\
School of Chemistry and Physics, University of Adelaide, Adelaide SA 5005, Australia}
\address[ITEP]{Institute for Theoretical and Experimental Physics, Moscow, 117218, Russia}
\address[NBI]{The Niels Bohr Institute, University of Copenhagen, Copenhagen, Denmark}

\begin{abstract}
In $N=1$ supergravity the scalar potential of the hidden sector may have degenerate
supersymmetric (SUSY) and non-supersymmetric Minkowski vacua. In this case local SUSY
in the second supersymmetric Minkowski phase can be broken dynamically.
Assuming that such a second phase and the phase associated with the physical
vacuum are exactly degenerate, we estimate the value of the cosmological constant.
We argue that the observed value of the dark energy density can be reproduced if
in the second vacuum local SUSY breaking is induced by gaugino condensation at a scale
which is just slightly lower than $\Lambda_{QCD}$ in the physical vacuum.
The presence of a third degenerate vacuum, in which local SUSY and electroweak (EW)
symmetry are broken near the Planck scale, may lead to small values of the quartic Higgs
self--coupling and the corresponding beta function at the Planck scale in the phase
in which we live.
\end{abstract}


\end{frontmatter}


\section{Introduction}
\label{intro}

It is commonly expected that the exploration of TeV scale physics at the LHC may
lead to the discovery of new physics phenomena beyond the Standard Model (SM)
that can shed light on the stabilisation of the EW scale. Indeed, if the SM is embedded
in a more fundamental theory characterized by a much larger energy scale (e.g. the
Planck scale $M_{Pl}\approx 10^{19}\,\mbox{GeV}$) than the EW scale, then due
to the quadratically divergent radiative corrections, the Higgs boson tends to acquire
a mass of the order of the larger energy scale; excessive fine-tuning is then required to
keep the Higgs mass around the observed value $\sim 125\,\mbox{GeV}$.

Despite the compelling arguments for physics beyond the SM, no signal or indication of
its presence has been detected at the LHC so far. Besides there are some reasons to believe
that the SM is extremely fine-tuned. Indeed, astrophysical and cosmological observations indicate that there
is a dark energy spread all over the Universe which constitutes $70\%-73\%$ of its energy density.
A fit to the recent data shows that its value is
$\rho_{\Lambda} \sim 10^{-123}M_{Pl}^4 \sim 10^{-55} M_Z^4$ \cite{Bennett:2003bz, Spergel:2003cb}.
At the same time much larger contributions should come from electroweak symmetry breaking
($\sim 10^{-67}M_{Pl}^4$) and QCD condensates ($\sim 10^{-79}M_{Pl}^4$). The contribution
of zero--modes is expected to push the vacuum energy density even higher up to $\sim M_{Pl}^4$, i.e.
\begin{equation}
\rho_{\Lambda}\simeq \sum_{bosons}\frac{\omega_b}{2}
- \sum_{fermions}\frac{\omega_f}{2}
\label{01}
\end{equation}
$$
=\int_{0}^{\Omega}\biggl[\sum_b\sqrt{|\vec{k}|^2+m_b^2}
-\sum_f\sqrt{|\vec{k}|^2+m_f^2}\biggr]\frac{d^3\vec{k}}{2(2\pi)^3}\sim -\Omega^4\,,
$$
where the $m_b$ and $m_f$ are the masses of bosons and fermions while $\Omega\sim M_{Pl}$.
Because of the cancellation needed between the contributions of different condensates to $\rho_{\Lambda}$,
the smallness of the cosmological constant should be regarded as a fine--tuning problem.

Here, instead of trying to alleviate fine-tuning, we postulate the exact degeneracy of different vacua.
The presence of such degenerate vacua was predicted by the so-called Multiple Point
Principle (MPP) \cite{Bennett:1994yx, Bennett:1993pj}, according to which Nature chooses values
of coupling constants such that many phases of the underlying theory should coexist. This scenario
corresponds to a special (multiple) point on the phase diagram of the theory where these phases
meet. The vacuum energy densities of these different phases are degenerate at the multiple point.

The MPP applied to the SM implies that the Higgs effective potential which is given by
\begin{equation}
V_{eff}(H) = m^2(\phi) H^{\dagger} H + \lambda(\phi) (H^{\dagger} H)^2\,,
\label{1}
\end{equation}
where $H$ is a Higgs doublet and $\phi$ is a norm of the Higgs field, i.e. $\phi^2=H^{\dagger} H$,
has two rings of minima in the Mexican hat with the same vacuum energy density \cite{Froggatt:1995rt}.
The radius of the little ring equals the EW vacuum expectation value (VEV) of the Higgs field,
whereas the second vacuum is taken to be at the Planck scale. The degeneracy of these vacua
can be achieved only if
\begin{equation}
\lambda(M_{Pl})\simeq 0 \,, \qquad\quad \beta_{\lambda}(M_{Pl})\simeq 0\,,
\label{2}
\end{equation}
where $\beta_{\lambda}=\frac{d \lambda(\phi)}{d \log\phi}$ is the beta--function of
$\lambda(\phi)$, which depends on $\lambda(\phi)$ itself, gauge $g_i(\phi)$ and top quark
Yukawa $g_t(\phi)$ couplings. Using MPP conditions (\ref{2}) the values of
the top quark and Higgs masses
were computed \cite{Froggatt:1995rt}
\begin{equation}
M_t=173\pm 5\,\mbox{GeV}\, ,\qquad M_H=135\pm 9\, \mbox{GeV}\,.
\label{21}
\end{equation}
The value of the Higgs mass specified above basically coincides with the lower bound on
the Higgs mass in the SM that comes from the vacuum stability constraint.
In previous papers the application of the MPP to the two Higgs doublet extension of the SM was considered
\cite{Froggatt:2006zc, Froggatt:2007qp, Froggatt:2008am}.
In particular, it was argued that the MPP can be used as a mechanism for the suppression of the flavour
changing neutral current and CP--violation effects \cite{Froggatt:2008am}.

Of critical importance here is the observation that the mass of the Higgs boson discovered at the LHC
is very close to the theoretical lower bound on $M_H$ in the SM mentioned above.
Thus the parameters of the SM can be extrapolated all the way up to $M_{Pl}$
without any inconsistency. Recently, using the extrapolation of the SM parameters up to the Planck
scale with full 3--loop RGE precision, it has been shown that (see \cite{Buttazzo:2013uya})
\begin{equation}
\lambda(M_{Pl}) =  -0.0143-0.0066\left(\frac{M_t}{\mbox{GeV}} - 173.34 \right)
\label{22}
\end{equation}
$$
+ 0.0018\left(\frac{\alpha_3(M_Z)-0.1184}{0.0007}\right) +0.0029\left(\frac{M_H}{\mbox{GeV}} -125.15 \right)\,.
$$
The computed value of  $\beta_{\lambda}(M_{Pl})$ also tends to be very small, so that the MPP
conditions (\ref{2}) are basically satisfied.

The success of the MPP in predicting the Higgs mass \cite{Froggatt:1995rt} suggests that we might also
use it for explaining the extremely low value of the dark energy density. In principle the smallness
of this energy density could be related to an almost exact symmetry. However at this moment
none of the available generalizations of the SM provides a satisfactory explanation for the smallness
of the cosmological constant. An exact global supersymmetry (SUSY) ensures zero value for the
energy density at the minimum of the potential of the scalar fields. Since superpartners of quarks
and leptons have not been observed yet, supersymmetry must be broken. In general the breakdown
of SUSY induces a huge and positive contribution to the total vacuum energy density of order $M_{S}^4$,
where $M_{S}$ is the SUSY breaking scale. The non--observation of superpartners of quarks and
leptons implies that $M_{S}\gtrsim 1 \ \mbox{TeV}$.

Here the MPP assumption is adapted to models based on $(N=1)$ local supersymmetry --
supergravity (SUGRA), in order to provide an explanation for the small deviation of the cosmological
constant from zero.

\section{Dark energy density in SUGRA models}
\label{sec-2}

The full $(N=1)$ SUGRA Lagrangian is specified
in terms of an analytic gauge kinetic function $f_a(\phi_{M})$ and
a real gauge-invariant K$a$hler function
$G(\phi_{M},\phi_{M}^{*})$, which depend on the chiral superfields
$\phi_M$. The function $f_{a}(\phi_M)$ determines the kinetic
terms for the fields in the vector supermultiplets and the gauge
coupling constants $Re f_a(\phi_M)=1/g_a^2$, where the index $a$
designates different gauge groups. The K$a$hler function is
a combination of two functions
\begin{equation}
G(\phi_{M},\phi_{M}^{*})=K(\phi_{M},\phi_{M}^{*})+\ln|W(\phi_M)|^2\,,
\label{3}
\end{equation}
where $K(\phi_{M},\phi_{M}^{*})$ is the
K$a$hler potential whose second derivatives define the
kinetic terms for the fields in the chiral supermultiplets.
$W(\phi_M)$ is the complete analytic superpotential of the
SUGRA model. Here we shall use standard supergravity
mass units: $\frac{M_{Pl}}{\sqrt{8\pi}}=1$.

The SUGRA scalar potential can be presented as a sum of $F$-- and
$D$--terms
\begin{equation}
V_{SUGRA}(\phi_M, \phi^{*}_M)=V_{F}(\phi_M, \phi^{*}_M)+V_{D}(\phi_M, \phi^{*}_M)\,,
\label{30}
\end{equation}
where the $F$-- and $D$--parts are given by
\begin{equation}
\begin{array}{rcl}
V_{F}(\phi_M,\phi^{*}_M)&=&\sum_{M,\,\bar{N}} e^{G}\left(G_{M}G^{M\bar{N}}
G_{\bar{N}}-3\right)\, ,\\[2mm]
V_{D}(\phi_M,\phi^{*}_M)&=&\frac{1}{2}\sum_{a}(D^{a})^2\,,\\[2mm]
D^{a}&=&g_{a}\sum_{i,\,j}\left(G_i T^a_{ij}\phi_j\right)\,,
\end{array}
\label{31}
\end{equation}
$$
G_M \equiv\partial_{M} G\equiv\partial G/\partial \phi_M\,, \quad
G_{\bar{M}}\equiv \partial_{\bar{M}}G\equiv\partial G/ \partial \phi^{*}_M\, .
$$
In Eq.~(\ref{31}) $g_a$ is the gauge coupling constant associated with
the generator $T^a$ of the gauge transformations.
The matrix $G^{M\bar{N}}$ is the inverse of the K$a$hler
metric $K_{\bar{N}M}$, i.e.
$$
G_{\bar{N}M}\equiv\partial_{\bar{N}}\partial_{M}G=
\partial_{\bar{N}}\partial_{M}K\equiv
K_{\bar{N}M}\,.
$$

In order to break supersymmetry in $(N=1)$ SUGRA models,
a hidden sector is introduced.
It contains superfields $z_i$, which are singlets under the SM
$SU(3)_C\times SU(2)_W\times U(1)_Y$ gauge group.
It is assumed that the superfields of the hidden sector interact
with the observable ones only by means of gravity.
If, at the minimum of the scalar potential, hidden sector fields
acquire VEVs so that at least one of their auxiliary fields
\begin{equation}
F^{M}=e^{G/2}G^{M\bar{P}}G_{\bar{P}}
\label{32}
\end{equation}
is non-vanishing, then local SUSY is spontaneously broken. At the same time a
massless fermion with spin $1/2$ -- the goldstino, which is a combination of
the fermionic partners of the hidden sector fields giving rise to the breaking
of SUGRA, is swallowed up by the gravitino which thereby becomes massive
$m_{3/2}=<e^{G/2}>$. This phenomenon is called the super-Higgs effect.

Usually the vacuum energy density at the minimum of the SUGRA scalar potential
(\ref{30})--(\ref{31}) tends to be negative. To show this, let us suppose that,
the function $G(\phi_{M},\phi_{M}^{*})$ has a stationary point, where all
derivatives $G_M=0$. Then it is easy to check that this point is
also an extremum of the SUGRA scalar potential. In the vicinity of
this point local supersymmetry remains intact while the energy
density is $-3<e^{G}>$. It implies that the vacuum energy density must
be less than or equal to this value. Therefore, in general, an enormous
fine--tuning must be imposed, in order to keep the total vacuum energy
density in SUGRA models around the observed value of the cosmological
constant.

\section{MPP inspired SUGRA models}
\label{sec-3}

The successful implementation of the MPP in $(N=1)$ supergravity requires
us to assume the existence of a vacuum in which the low--energy limit of the
considered theory is described by a pure supersymmetric model in flat Minkowski space
\cite{Froggatt:2003jm, Froggatt:2004gc, Froggatt:2005nb, Froggatt:2007qs, Froggatt:2011fc, Froggatt:2014jza}.
According to the MPP this vacuum and the physical one must be degenerate. Since the vacuum
energy density of supersymmetric states in flat Minkowski space is just zero,
the cosmological constant problem is thereby solved to first approximation.
Such a second vacuum is realised only if the SUGRA scalar potential has a minimum
where $m_{3/2}=0$. The corresponding minimum is achieved when
the superpotential $W$ for the hidden sector and its derivatives
vanish \cite{Froggatt:2003jm}, i.e.
\begin{equation}
W(z_m^{(2)})=0\,,
\label{33}
\end{equation}
\begin{equation}
\frac{\partial W(z_i)}{\partial z_m}\Biggl|_{z_m=z_m^{(2)}}=0
\label{34}
\end{equation}
where $z_m^{(2)}$ denote VEVs of the hidden sector
fields in the second vacuum.

The simplest K$a$hler potential and superpotential that satisfy
conditions (\ref{33}) and (\ref{34}) can be written as \cite{Froggatt:2003jm}
\begin{equation}
K(z,\,z^{*})=|z|^2\,,\qquad\qquad W(z)=m_0(z+\beta)^2\,.
\label{51}
\end{equation}
The hidden sector of this SUGRA model contains only one singlet
superfield $z$. If the parameter $\beta=\beta_0=-\sqrt{3}+2\sqrt{2}$,
the corresponding SUGRA scalar potential possesses two degenerate
minima with zero energy density at the classical level. One of them is a
supersymmetric Minkowski minimum that corresponds to $z^{(2)}=-\beta$.
In the other minimum of the SUGRA scalar potential
($z^{(1)}=\sqrt{3}-\sqrt{2}$) local supersymmetry is broken; so it can be
associated with the physical vacuum. Varying the parameter $\beta$ around
$\beta_0$ one can obtain a positive or a negative contribution from the
hidden sector to the total energy density of the physical vacuum. Thus $\beta$
can be fine--tuned so that the physical and second vacua are degenerate.

In general Eq.~(\ref{33}) represents the extra fine-tuning associated
with the presence of the supersymmetric Minkowski vacuum.
This fine-tuning can be to some extent alleviated in the no--scale inspired SUGRA models
with broken dilatation invariance \cite{Froggatt:2004gc, Froggatt:2005nb, Froggatt:2007qs}.
Let us consider a model with two hidden sector supermultiplets $T$ and $z$.
These superfields transform differently under the imaginary translations
($T\to T+i\beta,\, z\to z$) and dilatations ($T\to\alpha^2 T,\, z\to\alpha\,z$).
If the superpotential and K$a$hler potential of the hidden sector
of the SUGRA model under consideration are given by
\begin{equation}
\begin{array}{rcl}
K(T,\,z)&=&-3\ln\biggl[T+\overline{T}-|z|^2\biggr]\,,\\[2mm]
W(z)&=&\kappa\biggl(z^3+ \mu_0 z^2 \biggr)\,,
\end{array}
\label{61}
\end{equation}
then the corresponding tree level scalar potential of the hidden sector
is positive definite
\begin{equation}
V(T,\, z)=\frac{1}{3(T+\overline{T}-|z|^2)^2}
\biggl|\frac{\partial W(z)}{\partial z}\biggr|^2\,,
\label{6}
\end{equation}
so that the vacuum energy density vanishes near its global minima.
The scalar potential (\ref{6}) possesses two minima  at $z=0$ and
$z=-\frac{2\mu_0}{3}$ that correspond to the stationary points of
the hidden sector superpotential. In the first vacuum, where
$z=-\frac{2\mu_0}{3}$, local supersymmetry is broken so that
the gravitino becomes massive
\begin{equation}
m_{3/2}=\frac{4\kappa\mu_0^3}{27\biggl<\biggl(T+\overline{T}
-\frac{4\mu_0^2}{9}\biggr)^{3/2}\biggr>}\,.
\label{62}
\end{equation}
and all scalar particles get non--zero masses. Since one can
expect that $\mu_0\lesssim M_{Pl}$ and $\kappa\lesssim 1$,
SUSY is broken in this vacuum near the Planck scale. In the second
minimum, with $z=0$, the superpotential of the hidden sector vanishes
and local SUSY remains intact, so that the low--energy limit of
this theory is described by a pure SUSY model in flat Minkowski
space.

Of course, the inclusion of perturbative and non--perturbative corrections to the
Lagrangian of the no--scale inspired SUGRA model, which should depend on the
structure of the underlying theory, are expected to spoil the degeneracy of vacua
inducing a huge energy density in the vacuum where SUSY is broken. Moreover
in this SUGRA model the mechanism for the stabilization of the VEV
of the hidden sector field $T$ remains unclear. The model discussed above
should therefore be considered as a toy example only. This SUGRA model demonstrates that,
in $(N=1)$ supergravity, there might be a mechanism which ensures the vanishing
of vacuum energy density in the physical vacuum. This mechanism may also lead to
a set of degenerate vacua with broken and unbroken supersymmetry, resulting in
the realization of the multiple point principle.

\section{Estimation of the dark energy density}
\label{sec-4}

Let us now assume that a phenomenologically viable SUGRA model with degenerate vacua
of the type discussed in the previous section is realised in Nature. This implies that there are at
least two vacua which are exactly degenerate. In the first (physical) vacuum the spontaneous
breakdown of SUSY takes place near the Planck scale. We shall assume that in the second vacuum
SUSY remains intact and only vector supermultiplets, which correspond to the unbroken gauge
symmetry in the hidden sector, remain massless. These vector supermultiplets can give rise
to the breakdown of SUSY in the second vacuum which is caused by the formation of a
gaugino condensate induced in the hidden sector at the scale $\Lambda_{SQCD}\ll M_{Pl}$.

In this context it is worth to recall that the gaugino condensate does not actually break global
SUSY. Nonetheless, we can have a non-trivial dependence of the gauge kinetic function
$f_{X}(z_m)$ on the hidden sector superfields $z_m$. Such a dependence leads to auxiliary fields
corresponding to the hidden fields $z_m$
\begin{equation}
F^{z_m}\propto \frac{\partial f_X(z_k)}{\partial z_m}\bar{\lambda}_a\lambda_a+...
\label{16}
\end{equation}
acquiring non--zero VEVs, which are set by $<\bar{\lambda}_a\lambda_a>\simeq \Lambda_{SQCD}^3$.
Thus it is only via the effect of a non-renormalisable term that this gaugino condensate causes the SUSY breaking.
As a consequence the SUSY breaking scale $M_S^2 \sim \frac{\Lambda^3_{SQCD}}{M_{Pl}}$
is many orders of magnitude lower than $\Lambda_{SQCD}$ \cite{Nilles:1990zd}. Therefore the formation of a gaugino
condensate gives rise to a vacuum energy density
\begin{equation}
\rho^{(2)}_{\Lambda} \sim M_S^4 \sim \frac{\Lambda_{SQCD}^6}{M_{Pl}^2}\, .
\label{17}
\end{equation}

\begin{figure}
\begin{center}
{\includegraphics[width=0.49\textwidth]{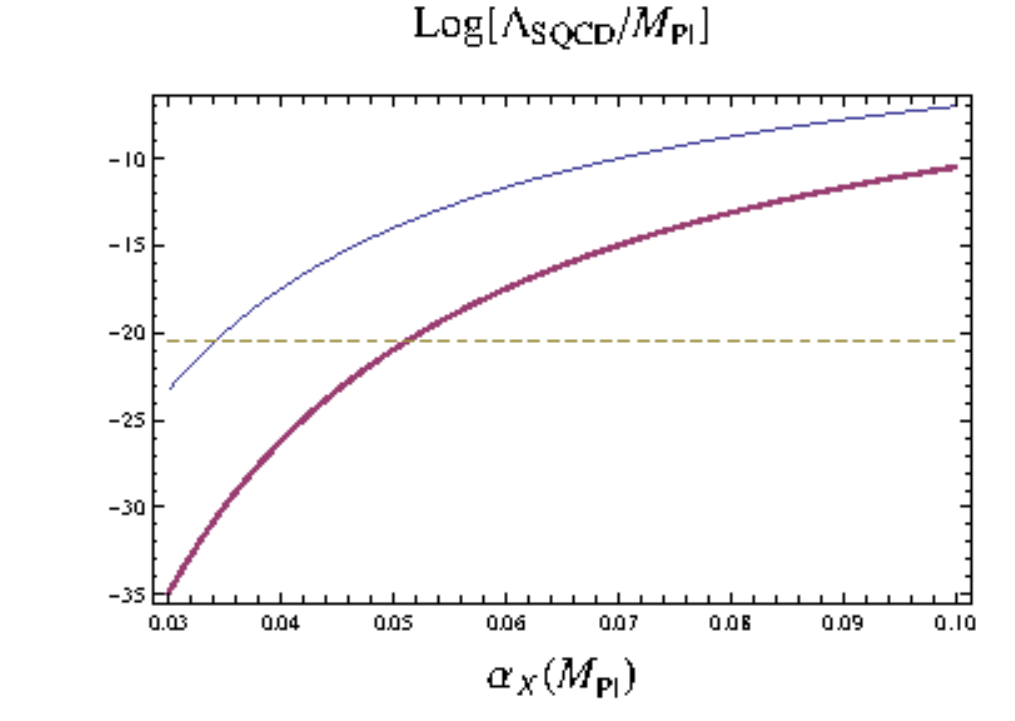}}
\end{center}
\caption{The value of $\log\left[\Lambda_{SQCD}/M_{Pl}\right]$ versus $\alpha_X(M_{Pl})$.
The thin and thick solid lines correspond to the $SU(3)$ and $SU(2)$ gauge symmetries, respectively.
The horizontal line is associated with the value of $\Lambda_{SQCD}$ that leads to the observed
value of the cosmological constant.}
\label{new-fig1}
\end{figure}

The postulated exact degeneracy of vacua then requires that the physical vacuum, in which
SUSY is broken near the Planck scale, has the same energy density as the phase where local
supersymmetry breakdown is induced by the gaugino condensate. From Eq.~(\ref{17})
it follows that the observed value of the cosmological constant can be reproduced if
$\Lambda_{SQCD}$ is relatively close to $\Lambda_{QCD}$ in the physical vacuum
\cite{Froggatt:2014jza}, i.e.
\begin{equation}
\Lambda_{SQCD}\sim \Lambda_{QCD}/10\,.
\label{18}
\end{equation}
Although there is no compelling theoretical reason to expect a priori that the two scales
$\Lambda_{SQCD}$ and $\Lambda_{QCD}$ should be relatively close or related,
$\Lambda_{QCD}$ and $M_{Pl}$ can be considered as the two most natural choices
for the scale of dimensional transmutation in the hidden sector.

For each given value of the gauge coupling $\alpha_X(M_{Pl})$ of the hidden sector gauge
interactions one can estimate the energy scale, $\Lambda_{SQCD}$, where the supersymmetric
QCD-like interactions become strong in the second vacuum. The analytical solution of the
one--loop renormalization group equation for $\alpha_X(Q)$ is given by
\begin{equation}
\frac{1}{\alpha_X(Q)}=\frac{1}{\alpha_X(M_{Pl})}+\frac{b_X}{4\pi}\ln\frac{M^2_{Pl}}{Q^2}\,,
\label{19}
\end{equation}
where $b_X$ is the one--loop beta function of the hidden sector gauge interactions. In particular,
$b_X=-9$ and $-6$ for the $SU(3)$ and $SU(2)$ gauge groups respectively.
Setting $\frac{1}{\alpha_X(\Lambda_{SQCD})}\to 0$ one finds
\begin{equation}
\Lambda_{SQCD}=M_{Pl}\exp\left[{\frac{2\pi}{b_X \alpha_X(M_{Pl})}}\right]\,.
\label{20}
\end{equation}
The dependence of $\Lambda_{SQCD}$ on $\alpha_X(M_{Pl})$ is shown in Fig.~1.
From Fig.~1 it follows that the value of $\Lambda_{SQCD}$ diminishes with decreasing
$\alpha_X(M_{Pl})$. The measured value of the dark energy density is reproduced when
$\alpha_X(M_{Pl})\simeq 0.051$ in the case of the model based on the $SU(2)$ gauge group and
$\alpha_X(M_{Pl})\simeq 0.034$ in the case of the $SU(3)$ SUSY gluodynamics. These values
of $\alpha_X(M_{Pl})$ correspond to $g_X(M_{Pl})\simeq 0.801$ and $g_X(M_{Pl})\simeq 0.654$
respectively. Thus in the case of the $SU(3)$ model the gauge coupling $g_X(M_{Pl})$
is just slightly larger than the value of the QCD gauge coupling at the Planck scale, i.e.
$g_3(M_{Pl})=0.487$ (see Ref.~\cite{Buttazzo:2013uya}), in the physical vacuum where we live.

\section{Implications for Higgs phenomenology}
\label{sec-5}

Now we shall discuss the possible implications of SUGRA models with degenerate vacua
for Higgs phenomenology. The presence of two degenerate vacua does not rule out the
possibility that there can exist another vacuum with the same energy density where
EW symmetry is broken near the Planck scale. Since in this third vacuum the Higgs VEV
is somewhat close to $M_{Pl}$ one must consider the interaction of the Higgs and hidden
sector fields. Thus the full scalar potential can be written:
\begin{equation}
V=V_{hid}(z_m) + V_0(H) + V_{int}(H, z_m)+...\,,
\label{21}
\end{equation}
where $V_{hid}(z_m)$ is the part of the scalar potential associated with the hidden sector,
$V_0(H)$ is the part of the full scalar potential that depends on the Higgs field only and
$V_{int}(H, z_m)$ corresponds to the interaction of the SM Higgs field with the hidden sector fields.
Here we assume that in the observable sector only one Higgs doublet acquires
a non--zero VEV and all other observable fields can be ignored in the first approximation.

Let us first consider the limit when $V_{int}(H, z_m)\to 0$. In this case, the Planck scale
VEV of the Higgs field would not lead to substantial variations of the VEVs of hidden sector fields.
Because of this the gauge couplings and $\lambda(M_{Pl})$ in the third and physical vacua
are expected to be approximately equal. Then the requirement of the degeneracy of all three vacua
should lead to the conditions (\ref{2}).

It is worth noting that in general in the vacuum with the Planck scale VEV of the Higgs doublet
the VEVs of the hidden sector fields should be very different from those in the physical
vacuum when $V_{int}(H, z_m)$ is not vanishingly small. Therefore in the third vacuum
the gauge couplings at the Planck scale, $\lambda(M_{Pl})$ and $m^2(M_{Pl})$, that
in general depend on the VEVs of the hidden sector fields, are not the same as in the physical
vacuum.

Nonetheless the interactions between the SM Higgs doublet and the hidden sector
fields can be rather weak near the third vacuum, i.e. $V_{int}(H, z_m)\ll M_{Pl}^4$,
if the VEV of the Higgs field is considerably smaller than $M_{Pl}$ (say $<H>\sim M_{Pl}/10$)
and the couplings of the SM Higgs doublet to the hidden sector fields are suppressed.
Then the VEVs of the hidden sector fields in the third and physical vacua can be basically identical.
As a result the gauge couplings and $\lambda(M_{Pl})$ in the third vacuum remain almost
the same as in the physical vacuum. On the other hand, the absolute value of $m^2$ in the Higgs
effective potential should  - although fixed to be small at the weak scale according
to  experiment - be much larger in the third vacuum. Indeed, in the
physical vacuum $|m^2|$ can be small because of the cancellation of different
contributions. However, in this case even small variations of the VEVs of the hidden sector fields
should spoil such cancellations. Although in the third vacuum $|m^2|$ is expected to be
many orders of magnitude larger than the EW scale, it can still be substantially
smaller than $M_{Pl}^2$ and $\langle H^{\dagger} H\rangle$ if the interactions between the
SM Higgs doublet and hidden sector fields are weak. In this limit the value of $V_{hid}(z_m)$
in the third vacuum remains almost the same as in the physical vacuum where
$V_{hid}(z^{(1)}_m)\ll M_{Pl}^4$. This means that the requirement of the degeneracy of vacua implies
that in the third vacuum $\lambda(M_{Pl})$ and $\beta_{\lambda}(M_{Pl})$ are approximately zero.
Since in this case the couplings in the third and physical vacua are basically identical,
the presence of such a third vacuum results in the predictions (\ref{2}) for $\lambda(M_{Pl})$ and
$\beta_{\lambda}(M_{Pl})$ in the physical vacuum.

\section{Conclusions}
\label{concl}

In $N=1$ supergravity (SUGRA)  supersymmetric (SUSY) and non-supersymmetric
Minkowski vacua originating in the hidden sector can be degenerate. This allows for
consistent implementation of the multiple point principle (MPP) which implies that
such vacua have the same vacuum energy densities. We presented SUGRA models
where the MPP assumption is realised. In the supersymmetric phase in flat Minkowski
space SUSY may be broken dynamically inducing a tiny vacuum energy density which
can be assigned, by virtue of MPP, to all other phases including the one in which we live.
We have argued that SUGRA models with degenerate vacua can lead to the measured value
of the dark energy density, as well as small values of $\lambda(M_{Pl})$ and $\beta_{\lambda}(M_{Pl})$.
This is realised in a scenario where the existence of at least three exactly degenerate vacua is postulated.
In the first (physical) vacuum SUSY is broken near the Planck scale and the small value of the
cosmological constant appears as a result of the fine-tuned precise cancellation of different
contributions. In the second vacuum the breakdown of local supersymmetry is induced by gaugino
condensation, which is formed at a scale which is slightly lower than $\Lambda_{QCD}$ in the physical
vacuum. In the third vacuum local SUSY and EW symmetry are broken near the Planck scale.

It is worth noting that our estimate of the tiny value of the cosmological constant
makes sense only if the vacua mentioned above are degenerate to very high accuracy.
In principle, a set of approximately degenerate vacua can arise if the underlying theory allows only
vacua which have a similar order of magnitude of space-time 4-volumes at the final stage of the
evolution of the Universe (This may imply the possibility of violation of the principle that the
future can have no influence on the past \cite{Bennett:1994yx}.). Since the sizes of these volumes
are determined by the expansion rates of the corresponding vacua associated with them,
only vacua with similar order of magnitude of dark energy densities are allowed.
Thus all vacua are degenerate to the accuracy of the value of the cosmological
constant in the physical vacuum.

\section{Acknowledgements}
\label{ack}

This work was supported by the University of Adelaide and the Australian Research Council through the ARC
Center of Excellence in Particle Physics at the Terascale and through grant LFO 99 2247 (AWT). HBN thanks
the Niels Bohr Institute for his emeritus status. CDF thanks Glasgow University and the Niels Bohr Institute
for hospitality and support.




\nocite{*}
\bibliographystyle{elsarticle-num}
\bibliography{cc-ichep2014}

\begin{thebibliography}{10}
\expandafter\ifx\csname url\endcsname\relax
  \def\url#1{\texttt{#1}}\fi
\expandafter\ifx\csname urlprefix\endcsname\relax\def\urlprefix{URL }\fi
\expandafter\ifx\csname href\endcsname\relax
  \def\href#1#2{#2} \def\path#1{#1}\fi

\bibitem{Bennett:2003bz}
{WMAP Collaboration}, {First year Wilkinson Microwave Anisotropy Probe (WMAP)
  observations: Preliminary maps and basic results}, Astrophys. J. Suppl. 148
  (2003) 1, [astro-ph/0302207].

\bibitem{Spergel:2003cb}
{WMAP Collaboration}, {First year {Wilkinson Microwave Anisotropy Probe (WMAP)}
  observations: Determination of cosmological parameters}, Astrophys. J. Suppl.
  148 (2003) 175, [astro-ph/0302209].

\bibitem{Bennett:1994yx}
D.~L. Bennett, C.~D. Froggatt, H.~B. Nielsen, {Nonlocality as an explanation
  for fine tuning and field replication in nature}, [hep-ph/9504294].

\bibitem{Bennett:1993pj}
D.~L. Bennett, H.~B. Nielsen, {Predictions for nonAbelian fine structure
  constants from multicriticality}, Int. J. Mod. Phys. A 9 (1994) 5155,
  [hep-ph/9311321].

\bibitem{Froggatt:1995rt}
C.~D. Froggatt, H.~B. Nielsen, {Standard model criticality prediction: Top mass
  $173 \pm 5\,\mbox{GeV}$ and Higgs mass $135 \pm 9\,\mbox{GeV}$}, Phys. Lett.
  B 368 (1996) 96, [hep-ph/9511371].

\bibitem{Froggatt:2006zc}
C.~D. Froggatt, L.~V. Laperashvili, R.~B. Nevzorov, H.~B. Nielsen, M.~Sher,
  {Implementation of the multiple point principle in the two-Higgs doublet
  model of type II}, Phys. Rev. D 73 (2006) 095005, [hep-ph/0602054].

\bibitem{Froggatt:2007qp}
C.~D. Froggatt, R.~Nevzorov, H.~B. Nielsen, D.~Thompson, {Fixed point scenario
  in the Two Higgs Doublet Model inspired by degenerate vacua}, Phys. Lett. B
  657 (2007) 95, [arXiv:0708.2903 [hep-ph]].

\bibitem{Froggatt:2008am}
C.~D. Froggatt, R.~Nevzorov, H.~B. Nielsen, D.~Thompson, {On the origin of
  approximate custodial symmetry in the Two-Higgs Doublet Model}, Int. J. Mod.
  Phys. A 24 (2009) 5587, [arXiv:0806.3190 [hep-ph]].

\bibitem{Buttazzo:2013uya}
D.~Buttazzo, G.~Degrassi, P.~P. Giardino, G.~F. Giudice, F.~Sala, A.~Salvio,
  A.~Strumia, {Investigating the near-criticality of the Higgs boson}, JHEP
  1312 (2013) 089, [arXiv:1307.3536 [hep-ph]].

\bibitem{Froggatt:2003jm}
C.~D. Froggatt, L.~V. Laperashvili, R.~Nevzorov, H.~B. Nielsen, {Cosmological
  constant in SUGRA models and the multiple point principle}, Phys. Atom. Nucl.
  67 (2004) 582, [hep-ph/0310127].

\bibitem{Froggatt:2004gc}
C.~D. Froggatt, L.~V. Laperashvili, R.~Nevzorov, H.~B. Nielsen, {No-scale
  supergravity and the multiple point principle}, hep-ph/0411273.

\bibitem{Froggatt:2005nb}
C.~D. Froggatt, R.~Nevzorov, H.~B. Nielsen, {On the smallness of the
  cosmological constant in SUGRA models}, Nucl. Phys. B 743 (2006) 133,
  [hep-ph/0511259].

\bibitem{Froggatt:2007qs}
C.~D. Froggatt, R.~Nevzorov, H.~B. Nielsen, {Smallness of the cosmological
  constant and the multiple point principle}, J. Phys. Conf. Ser. 110 (2008)
  072012, [arXiv:0708.2907 [hep-ph]].

\bibitem{Froggatt:2011fc}
C.~D. Froggatt, R.~Nevzorov, H.~B. Nielsen, {Dark Energy density in models with
  Split Supersymmetry and degenerate vacua}, Int. J. Mod. Phys. A 27 (2012)
  1250063, [arXiv:1103.2146 [hep-ph]].

\bibitem{Froggatt:2014jza}
C.~D. Froggatt, R.~Nevzorov, H.~B. Nielsen, A.~W. Thomas, {Cosmological
  constant in SUGRA models with Planck scale SUSY breaking and degenerate
  vacua}, Phys. Lett. B 737 (2014) 167, [arXiv:1403.1001 [hep-ph]].

\bibitem{Nilles:1990zd}
H.~P. Nilles, {Gaugino Condensation and Supersymmetry Breakdown}, Int. J. Mod.
  Phys. A 5 (1990) 4199.

\end{thebibliography}







\end{document}